\documentclass[preprint,aps]{revtex4}
\usepackage{graphicx}% Include figure files
\usepackage{dcolumn}% Align table columns on decimal point
\usepackage{bm}% bold math

\begin{document}
%\begin{frontmatter}
\title{Structural and magnetic properties of half-heusler alloys NiCrZ (Z = Si, P, Ge, As, Te): First principle study.}
\author{Van An Dinh} 
\email{divan@cmp.sanken.osaka-u.ac.jp.}
%\corauth[cor]{Corresponding author.}
\author{Kazunori Sato and Hiroshi Katayama-Yoshida}
\affiliation{The Institute of Scientific and Industrial Research, Osaka University, Mihogaoka 8-1, Ibaraki, Osaka 567-0047, Japan.}

\begin{abstract}
  We present a first principle study of new class of high-$T_c$ half-heusler ferromagnets NiCrZ (Z = Si, P, Ge, As, Te). The structure and magnetic properties are investigated through the calculation of the  electronic structure, equilibrium lattice constant, magnetic exchange interaction $J_{ij}$ and  Curie temperature $T_c$. The role of $sp$-elements and the influence of lattice expansion/compression are also studied. In alloys having 20 valence electrons, a pseudo-gap of the majority band can be formed at Fermi level. Otherwise, the half-metallicity and ferromagnetism at temperatures much higher than room temperature are found to be stable in a wide range of lattice expansion. Based on these results, NiCrZ can be expected to be promising materials for spintronics.
\end{abstract}
%
%
%\keywords{{\it ab initio} calculation, half-heusler, high-$T_c$, ferromagnetism, half-metallic, spintronics, materials design}
%
%\end{frontmatter}
%
%\section{Introduction}
\maketitle
 To exploit the great potential of spintronics, seeking for new magnetic materials and half metallic ferromagnets is one of the most important issues in materials physics. Half metallic ferromagnets, whose the electronic structure behaves like metals w.r.t the electrons of one spin direction and like semiconductors w.r.t. the spins in the opposite direction, have the extreme spin polarization at Fermi level $E_F$ and can achieve the ferromagnetism at temperatures higher than room temperature. 
 
 Since the pioneer work on the half-metallicity in half-heusler alloy NiMnSb by de Groot and collaborators \cite{groot} in 1983, heusler and half-heusler alloys have been attracted great attention in both theoretical and experimental study. Especially, the family of half-heusler alloy XYZ (X and Y are transition metals, and Z is an $sp-$valence element), 
 has been one of the most interested compounds \cite{webster,galanakis,han,soulen,bowen,nanda,an} due to the possibility of half-metallic ferromagnetism being achieved. Besides the extreme spin polarization has been verified  \cite{han,soulen,bowen}, the ferromagnetism in the half-Heusler NiMnZ at temperatures ranging between 500K and 730K has been reported for Z= Pd, Pt and Sb \cite{webster} by experiments. Theoretically, the first principle study has predicted the half-metallic character in their electronic structure of many half-heusler alloys such as FeMnSb, CoMnSb and FeCrSb, etc. \cite{galanakis,nanda}. Recently, the half-metallic ferromagnetism with very high Curie temperature ($T_c$) in a new class of half-heusler alloys NiMnZ (particularly, $T_c\approx 1000$K for NiMnSi) has been predicted \cite{an}. Also, the half-metallic ferromagnetism with $T_c$ higher than room temperature can be expected even if Cr replace Mn in NiMnZ.   
   
   In this Letter, based on the first principle study of the structural and magnetic properties of NiCrZ (Z = Si, P, Ge, As and Te) we propose a new class of high-$T_c$ half-metallic ferromagnets. For this aim, we have carried out the calculation as follows.  First, we perform the total energy calculation to evaluate the equilibrium lattice constant (ELC) by means of the generalized gradient approximation (GGA) within the framework of ultrasoft pseudo-potential (UPP) (thanks to STATE-Senri) and both of the muffin-tin (MTA) and atomic sphere approximations (ASA) within KKR-LSDA (thanks to AKaiKKR). Using ELC, we next calculate the density of states (DOS), magnetic exchange interaction $J_{ij}$ between magnetic sites.  Finally, using $J_{ij}$ as the input data we calculate $T_c$ by employing three statistical approaches: mean field approximation (MFA), random phase approximation (RPA) and Monte Carlo simulation (MC).   

Half-heusler alloys NiCrZ have $C1_b$ structure. The study of structural property  \cite{nanda,an} shows the most stable configuration of NiYZ should be in $\alpha$ phase (except Z being a lighter $sp$-element such as C and N)  with Ni being located in the octahedral coordinated pocket ($\frac{1}{4},\frac{1}{4},\frac{1}{4}$), Y at ($0,0,0$), an $sp$-element Z at the body center ($\frac{1}{2},\frac{1}{2},\frac{1}{2}$) and a vacant site at $(\frac{3}{4},\frac{3}{4},\frac{3}{4})$.  Ni atom possesses 10 valence electrons and Cr has 6 electrons. Therefore, the numbers $N_t$ of valence electrons per formula unit are 20, 21 and 22 corresponding to $sp$-elements of group IV, V and VI, respectively. Regarding to "rule of 18" of the half-heusler alloys $M=N_t-18$, the total magnetic moment $M$ of NiCrZ should be 2$\mu_B$, $3\mu_B$ and $4\mu_B$ corresponding to Z of the groups IV, V and VI, respectively.

\begin{table}
\begin{center}
\caption{ELC (au) by means of  GGA within UPP ($a_{UP}$), MTA ($a_{MT}$) and ASA ($a_{AS}$) within KKR-LSDA.}
%\vskip 0.2cm
\label{tab1}
\begin{tabular}{@{\hspace{\tabcolsep}%
				\extracolsep{\fill}}llll} \hline
Alloy &$a_{UP}$&$a_{AS}$&$a_{MT}$\\ 
 \hline
NiCrSi\.&10.2714 & 10.3243&10.4450\\\hline
NiCrP&\.10.2622&10.3823&10.5390\\\hline
NiCrGe\.&10.4521&10.5523&10.6652\\\hline
NiCrAs\.&10.5631&10.6310&10.7966\\\hline
NiCrTe\.&--&11.5116&11.7980\\\hline
\end{tabular}
\end{center}
\end{table}
 Table~\ref{tab1} shows ELC calculated by means of three approximations: GGA within framework of UPP, MTA and ASA within LSDA. As can be seen from Tab.~\ref{tab1}, MFA gives the lattice constant $a_{MT}$ larger than ASA and UPP, whereas UPP gives the smallest one. However, the deviation between $a_{UP}$, $a_{AS}$ and $a_{MT}$ is considerably small ($a_{AS}-a_{UP} \approx 1$\%
  for NiCrSi). In general, the lattice constant is larger if the atom radius is larger. However, UPP gives $a_{UP}$ of NiCrP smaller than NiCrSi. Referring to the case of NiMnZ  \cite{an}, we can also  expect that the real lattice constant of NiCrZ might be arranged in the range from $a_{MT}$ to $a_{UP}$.  
\begin{figure} 
\begin{center}
\includegraphics[width=\linewidth]{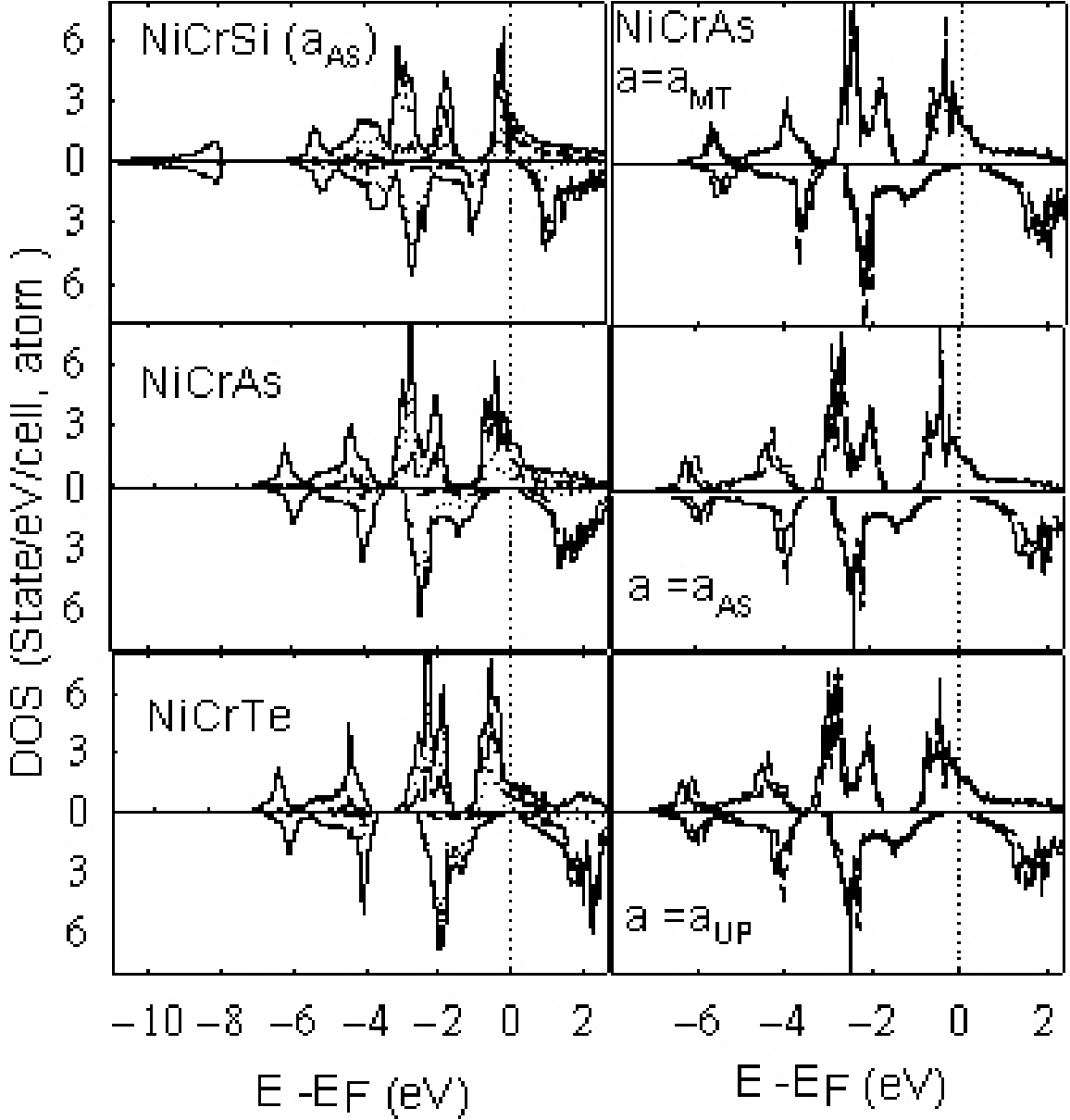}
\caption{DOS of NiCrSi, NiCrAs and NiCrTe at $a=a_{AS}$ are given in left panel. Volume dependence of DOS within LSDA and GGA is shown for NiCrAs in right panel .}
\label{DOSf} 
\end{center}
%\vskip -0.6cm
\end{figure}

The left panel of Fig.~\ref{DOSf} illustrates DOS of the typical alloys standing in the compounds which possess the different numbers of valence electrons. The DOS of three alloys possessing 20 valence electrons (NiCrSi), 21 valence electrons (NiCrAs) and 22 valence electrons (NiCrTe) are plotted. The change of the electronic structure w.r.t the lattice constant is shown in the right panel for NiCrAs as a representative case. In addition, it should be noted that NiCrZ alloys with Z = Sn, Sb, Se and S are metals at ELC. 
Furthermore, NiCrZ with Z being an element of group IV exhibits a pseudo-gap  of the majority band at $E_F$ in UPP calculation, and its electronic structure shows the low spin behavior at ELC but the high spin picture at sufficiently larger $a$ (about 0.5\%
 larger than ELC). 

  As seen from the left panel of Fig.~\ref{DOSf}, while almost states of $3d$ states of Ni concentrate at lower energies (in valence band) and are occupied, the $3d$ states of Cr are distributed at both of the lower and higher energies and strongly polarized. The  $3d-$majority spin states of Cr distribute at the lower energies and are almost occupied. These states hybridize with $3d$ states of Ni (and $3p$ of Z elements) to form a band at $E_F$. The minority spin states of Cr shift to the higher energies and are unoccupied, leading to the gap  (that $E_F$ falls into) being formed in the minority band and the half-metallic character. For alloys possessing 21 valence electrons, Mn-$3d$ states in the majority gap split into two bands $e_g$ and $t_a$, and $E_F$ lies at the vary narrow valley which is formed between these bands. Moreover, while the half-metallic character is obvious in DOS calculated by LSDA, GGA combined in UPP gives a very narrow pseudo-gap at $E_f$ in the majority band of NiCrGe and NiCrAs, and the half-metallic behavior might be disappeared. The calculated total magnetic moment $M$ of NiCrSi is exactly $2\mu_B$ in consistence with "rule of 18". Similarly, we have also obtained the half-metallic behavior of the remaining alloys NiCrZ (Z = P, Ge, As, Te) and the total magnetic moment of $2\mu_B$ for Z of group IV and $3\mu_B$ for Z of group V at $a = a_{UP}$, $a_{AS}$ and $a_{MT}$ (except NiCrGe and NiCrAs at $a=a_{MT}$. See also Tab.~\ref{tab2}), and $4\mu_B$ for Te at $a =  a_{AS}$ and $a_{MT}$ .
 
  To study the influence of the lattice expansion (or compression), we draw the DOS of NiCrAs at three different ELC in the right panel of Fig.~\ref{DOSf}. LSDA (black line) and GGA (blue line) calculations are given. Since the minority states of Cr-$3d$ electrons in GGA result is shifted toward the higher energies while the majority band is mostly kept unchanged, the minority gap in GGA is wider than LSDA. For NiCrAs, the half-metallicity is destroyed at $a_{MT}$. As the lattice constant $a$ decreases, $E_F$ shifts toward the conduction band, the half-metallicity becomes more stabilized; however, the half-metallicity can be destroyed if $a$ is smaller than the threshold value at which $E_F$ falls into the conduction band. Except NiCrAs and NiCrGe whose the half-metallicity might be destroyed at $a\ge a_{MT}$, our calculation shows that the proposed alloys preserve the half-metallicity in a wider range of the expansion (compression) of the lattice cells.
\begin{figure}
\begin{center}
\includegraphics[width=0.85\linewidth]{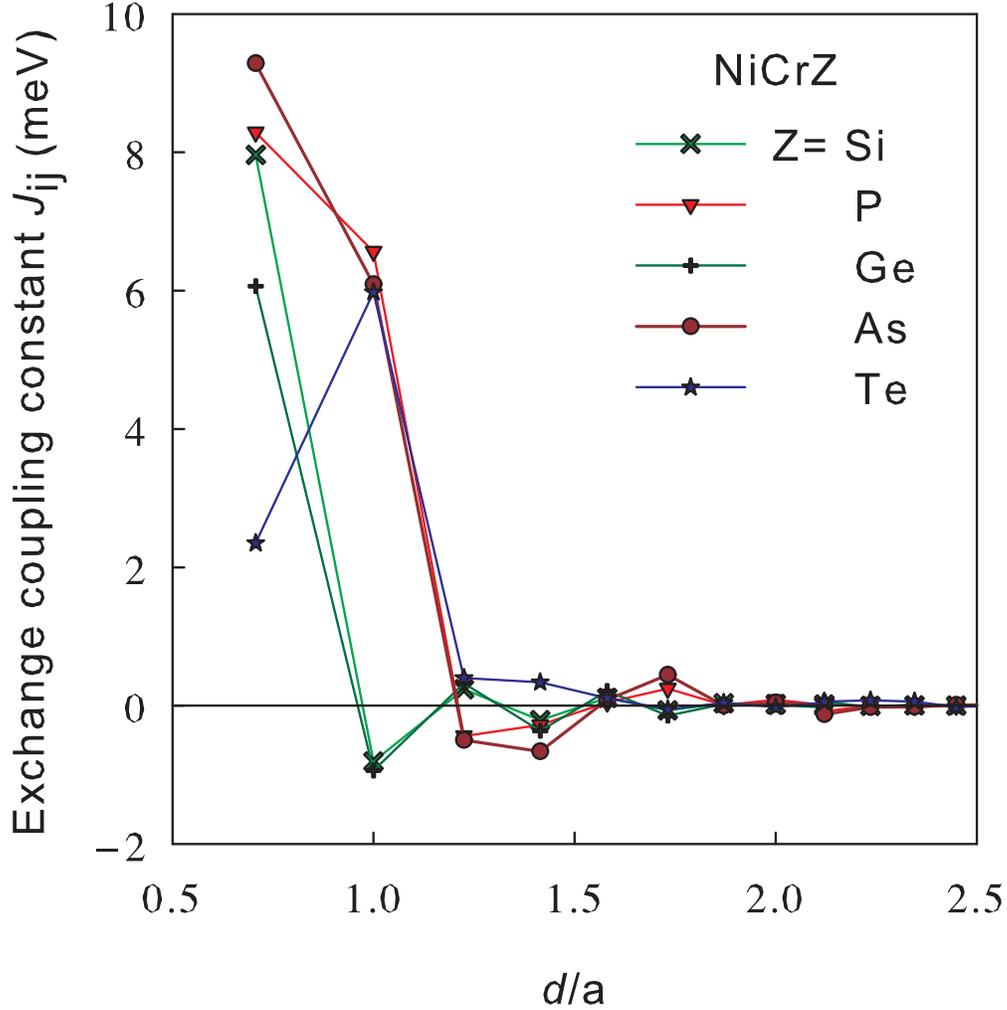}
\caption{Cr-Cr effective exchange coupling constant $J_{ij}$ vs. distance $d$ in units of lattice constants of NiCrZ. $J_{ij}$ is calculated at lattice constant $a=a_{AS}$. \label{jij}}
\end{center}
%\vskip -0.6cm
\end{figure}

%\begin{table}
%\begin{center}
%\caption{Distance dependence of magnetic exchange interaction. $+/-$ denotes FM/AF}
%\vskip 0.2cm
%\label{tab1}
%\begin{tabular}{@{\hspace{\tabcolsep}%
%				\extracolsep{\fill}}cccccc} \hline
%$N_t$&$1st$ &$2nd$&$3rd$&$4th$&$5th$\\ 
% \hline
%20&+&-&+&-&+\\\hline
%21&+&+&-&-&+\\\hline
%22&+&+&+&+&+\\\hline
%\end{tabular}
%\end{center}
%\vskip -0.5cm
%\end{table}
  To investigate the ferromagnetism in NiCrZ alloys, we calculate the magnetic exchange interaction $J_{ij}$ at three values of ELC. %Once $J_{ij}$ is obtained, the statistical methods such as MFA, RPA and MC are employed to calculate $T_c$.
 The exchange interaction $J_{ij}$ between two impurities at $ith$ and $jth$ sites, which are embedded in the ferromagnetic medium, is efficiently calculated by utilizing the magnetic force theorem. The frozen potential approximation  \cite{oswald} is employed and Liechtenstein formula  \cite{liech} is used for evaluation of $J_{ij}$. 
 %The total energy change due to the infinitesimal rotations of two magnetic moments at the $ith$ and $jth$ sites is calculated by using the magnetic force theorem, and the effective exchange interaction $J_{ij}$ is calculated via the mapping of the total energy change onto the classical Heisenberg model $H=-\Sigma_{i\ne j}{J_{ij}\mathbf{e}_i\mathbf{e}_j}$, where $\mathbf{e}_i$ denotes a unit vector parallel to the magnetic moment at the $ith$ site.
    $J_{ij}$ of NiCrZ at $a=a_{AS}$ is illustrated in Fig.\ref{jij}. For alloys with $N_t=20$, the magnetic exchange interaction $J_{ij}$ is ferromagnetic at the $1st$, $3rd$ and $5th$ nearest neighbors, but anti-ferromagnetic for the $2nd$ and $4th$ nearest neighbor pairs. For $N_t=21$, $J_{ij}$ is ferromagnetic for the $1st$, $2nd$ $5th$ and $6th$ nearest neighbor pairs. Similar to NiMnZ \cite{an}, $J_{ij}$ becomes ferromagnetic at distances up to the $5th$ nearest neighbors for $N_t=22$. Thus, $J_{0j}$ with $j=1$ and $j=2$ becomes larger with increasing $N_t$. For NiCrTe, due to the larger distance between atoms, $J_{01}$ becomes remarkably smaller than that of the other half-heusler alloys. %The similar trend is also seen in CMR manganites \cite{soloviev}.
   
\begin{figure}
\begin{center}
\includegraphics[width=\linewidth]{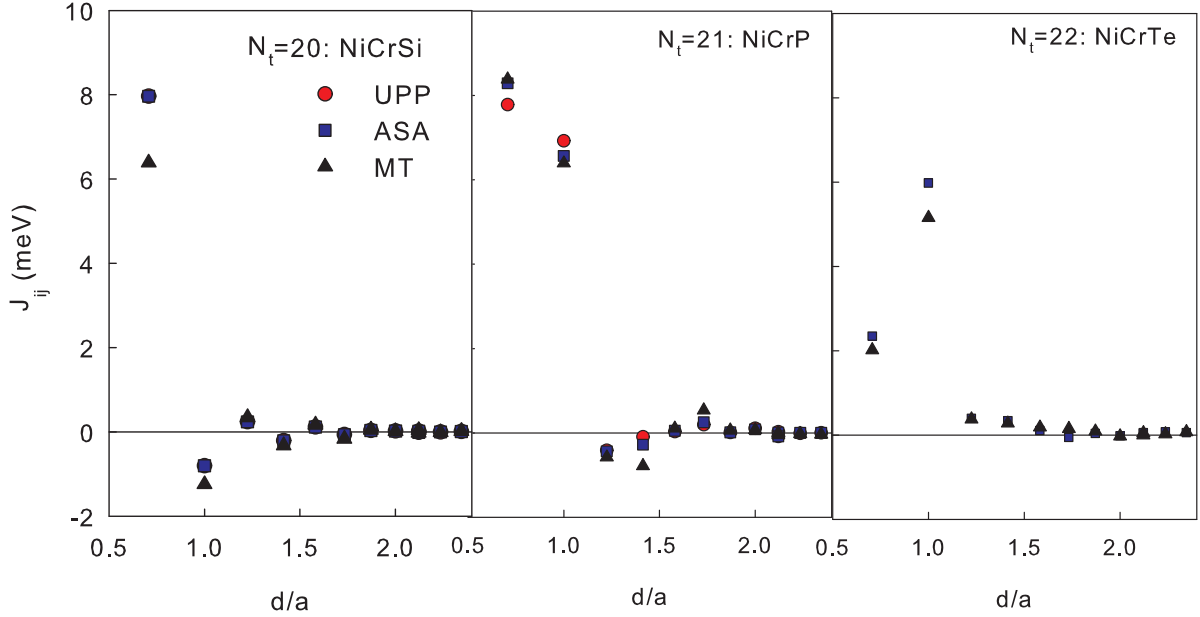}
\caption{$J_{ij}$ vs. lattice expansion. \label{jija}}
\end{center}
%\vskip -0.6cm
\end{figure}

   In order to study the volume expansion effect on the magnetic exchange interaction as well as $T_c$, we demonstrate the change of $J_{ij}$ versus the lattice expansion in Fig.~\ref{jija}. $J_{ij}$ of the typical half-heusler alloys corresponding to $N_t= 20, 21$ and $22$ is calculated for the nearest neighbor pairs at lattice constants $a_{UP}$ (circles), $a_{AS}$ (squares) and $a_{MT}$ (triangles). Generally, $J_{02}$ of all alloys decreases with increasing $a$. $J_{01}$ of alloys possessing 21 valence electrons increases with $a$, whereas $J_{01}$ of alloys with $N_t= 20 $ and $22$ decreases. Since the dominant contributions to $T_c$ come mostly from the nearest neighbors at short distances (say, at the $1st$ and $2nd$ neighbor pairs), $T_c$ of alloys with $N_t=20$ and 22 actually decreases with $a$. For NiCrP, $T_c$ can slightly increases in the range from $a_{UP}$ to $a_{AS}$ because the $J_{01}$ considerably increases, but since $J_{01}$'s change according to an increase of $a$ at the range of larger $a$ is very small whereas $J_{04}$ becomes more anti-ferromagnetic, $T_c$ decreases for $a> a_{AS}$. %This trend is also illustrated in Tab.~\ref{tab2}. 
    
\begin{table}
\begin{center}
\caption{Predicted $T_c$ (K) calculated by MFA, RPA and MC at $a = a_{MT}, a_{AS}$, $a_{UP}$.}
%\vskip 0.1cm
\label{tab2}
\begin{tabular}{@{\hspace{\tabcolsep}%
				\extracolsep{\fill}}llrrr} \hline
Alloy &$a$& $T_c^{MF}$& $T_c^{RP}$&$T_c^{MC}$\\ 
%&&(K)&(K)&(K)\\
 \hline
 NiCrSi&$a_{UP}$&778 & 606&634\\
($N_t=20$)&$a_{AS}$&758& 591&615\\
&$a_{MT}$&605&477&493\\\hline
  NiCrGe&$a_{UP}$&669 &562&550\\
($N_t=20$)&$a_{AS}$&587 &465&470\\\hline
%  &$a_{MT}$&232& 204& & Metal*\\
 NiCrP   &$a_{UP}$&974&777&805\\
($N_t=21$)&$a_{AS}$&992&784&810\\
  &$a_{MT}$&953&739&782\\\hline
NiCrAs&$a_{UP}$&1044 & 808&858\\
($N_t=21$)&$a_{AS}$&1025&785&829\\\hline
%  &$a_{MT}$&818& 449& 565 &Metal*\\
%  &$a_{UP}$&877& 705&736\\
NiCrTe &$a_{AS}$&665& 588&605\\ 
  ($N_t=22$)&$a_{MT}$&601& 544&562\\\hline
\end{tabular}
 \end{center}
\end{table}
 After calculating $J_{ij}$, the calculation of $T_c$ is carried out by three statistical approaches: the Monte Carlo simulation $T_c^{MC}$, the mean field approximation ($k_BT_c^{MF}=\frac{2}{3}\sum_{j}{J_{0j}}$, where $k_B$ is Boltzmann constant), and the random phase approximation \cite{hilbert,bouz} ($(k_BT_c^{RP})^{-1}=\frac{3}{2N}\sum_{\bf q}{(J(0)-J({\bf q}))^{-1}}$, where $J({\bf q})$ is the Fourier transform of the exchange parameter $J_{ij}$). Using $J_{ij}$, we evaluate $T_c$. To perform MC simulation, the Metropolis algorithm \cite{binder} is applied to calculate the thermal average of the magnetization $M$ and its powers. Then, the cumulant crossing method proposed by Binder  \cite{binder} is employed and the fourth order cumulant $U_4$ is calculated as a function of temperature for different cell sizes ($14\times 14\times 14, 16\times 16\times 16$, and $18\times 18\times 18$ conventional fcc cells) to find the universal fixed-point at $T_c$. Obtained results are shown in Tab.~\ref{tab2}. 
 
 It should be emphasized that with a similar method our theoretical $T_c^{MC}$ (745K) of NiMnSb \cite{an} is in very good agreement with the experimental result ($T_c=730$K)  \cite{webster}. Similar to the case of NiMnZ \cite{an}, MFA overestimates $T_c$ even at 100\%
  magnetic atoms, whereas RPA underestimates. As usual, the MC simulation gives the $T_c$ between values obtained by RPA and MFA at the same lattice constant. As expected, $T_c^{MC}$ of all alloys is higher than room temperature and arranges from 858K to 550K (at $a=a_{UP}$). The lowest $T_c$ corresponds to NiCrGe with $T_c=550$K at $a=a_{UP}$.  The highest $T_c$ ($858$K at $a_{UP}$) corresponds to NiCrAs and the next are NiCrP (805K), NiCrSi (634K) and NiCrTe (605K). It is noted that, obtained $T_c$ here is calculated based on the electronic structure within KKR-LSDA. GGA combined in UPP shows a pseudo-gap in the majority band of NiCrZ which possesses 20 valence electrons; hence, UPP calculation will shows a lower $T_c$ for NiCrGe and NiCrSi. Except NiCrAs and NiCrGe whose the half-metallicity is considerably sensitive with an increase of the lattice constant because of $E_F$ closed to the valence band edge, the half-metallicity in the remaining alloys can be preserved when $a$ varies in the range of $a_{UP}\le a\le a_{MT}$. 
 
In summary, we have investigated the structure and magnetic properties of the half-heusler alloys NiCrZ (Z = Si, P, Ge, As, Te). The phase stability is considered. The ELCs are predicted by three approximations within the pseudo-potential method and KKR-LSDA as well. Using the obtained equilibrium lattice constants, the magnetic exchange interaction is calculated and then the evaluation of $T_c$ is performed by employing MFA, RPA and Monte Carlo simulation. The volume dependence of the electronic structure and magnetic exchange interaction as well as $T_c$ are also discussed. Following the obtained results, we propose a new half-metallic high-$T_c$ half-heusler alloys NiCrZ (Z = Si, P, Ge, Te) whose half-metallicity can be preserved in a wide range of lattice expansion.

  {\bf Acknowledgements}
  
  This research was partially supported by a Grant-in-Aid for Scientific Research in Priority Areas "Quantum Simulators and Quantum Design" (No. 17064014) and "Semiconductor Nanospintronics," a Grand-in-Aid for Scientific Research for young researchers, JST- CREST, NEDO-nanotech, the 21st Century COE, and the JSPS core-to-core program "Computational Nano-materials Design." We are grateful to Prof. H. Akai and Prof Y. Morikawa (Osaka Univ.) for providing us with the first principle calculation packages.
  
%\makefigurecaptions
%\makefigurecaptions
\end{document}